\documentstyle[12pt,epsf]{article}
\begin{document}

\setlength{\topmargin}{0.1 in}
\setlength{\headheight}{0.1 in}
\setlength{\topskip}{0 in}
\setlength{\textheight}{8.5 in}
\setlength{\footskip}{0.5 in}

\def\F{{\cal F}}
\def\ETA{\zeta}

\def\op{{\cal O}}
\def\lsim{\mathrel{\lower4pt\hbox{$\sim$}}\hskip-12pt\raise1.6pt\hbox{$<$}\;
}
\def\Dd{\psi}
\def\pp{\lambda}
\def\ket{\rangle}
\def\BAR{\bar}
\def\xba{\bar}
\def\fa{{\cal A}}
\def\fm{{\cal M}}
\def\fl{{\cal L}}
\def\ufs{\Upsilon(5S)}
\def\gsim{\mathrel{\lower4pt\hbox{$\sim$}}
\hskip-10pt\raise1.6pt\hbox{$>$}\;}
\def\ufour{\Upsilon(4S)}
\def\xcp{X_{CP}}
\def\ynotcp{Y}
\vspace*{-.5in}
\def\ETAp{\ETA^\prime}
\def\bfb{{\bf B}}
\def\fd{r_D}
\def\fb{r_B}
\def\ed{\ETA_D}
\def\eb{\ETA_B}
\def\hatA{\hat A}
\def\hatfd{{\hat r}_D}
\def\hated{{\hat\ETA}_D}

\def\uglu{\hskip 0pt plus 1fil
minus 1fil} \def\uglux{\hskip 0pt plus .75fil minus .75fil}

\def\slashed#1{\setbox200=\hbox{$ #1 $}
    \hbox{\box200 \hskip -\wd200 \hbox to \wd200 {\uglu $/$ \uglux}}}
\def\slpar{\slashed\partial}
\def\sla{\slashed a}
\def\slb{\slashed b}
\def\slc{\slashed c}
\def\sld{\slashed d}
\def\sle{\slashed e}
\def\slf{\slashed f}
\def\slg{\slashed g}
\def\slh{\slashed h}
\def\sli{\slashed i}
\def\slj{\slashed j}
\def\slk{\slashed k}
\def\sll{\slashed l}
\def\slm{\slashed m}
\def\sln{\slashed n}
\def\slo{\slashed o}
\def\slp{\slashed p}
\def\slq{\slashed q}
\def\slr{\slashed r}
\def\sls{\slashed s}
\def\slt{\slashed t}
\def\slu{\slashed u}
\def\slv{\slashed v}
\def\slw{\slashed w}
\def\slx{\slashed x}
\def\sly{\slashed y}
\def\slz{\slashed z}

%%%%
%
\begin{flushright}
{AMES-HET 03-02}\\
{BNL-HET-03/8}\\
\end{flushright}
%
%%%%

\vspace*{1.2 in}

\begin{center}
{\large\bf
Role of Charm Factory in Extracting CKM-Phase Information
via $B\to DK$}
\end{center}

\vspace*{0.7 in}

\begin{center}
{David Atwood}\\
{Dept. of Physics and Astronomy, Iowa State University, Ames,
IA 50011}\\
\bigskip
\bigskip
{Amarjit Soni}\\
{Theory Group, Brookhaven National Laboratory, Upton, NY
11973}
\end{center}

\vspace*{1.0 in}

\begin{center}
{\bf Abstract}
\end{center}

In this paper we study the impact of data that can be obtained from a
Charm Factory on the determination of the 
CKM
parameter $\gamma$ from
decays of the form $B\to D^0 K$ where the $D^0$ decays to specific
inclusive and exclusive final states. In particular, for each exclusive
final state $f$, the charm factory can determine the strong phase
difference between $D^0\to f$ and $\xba D^0\to f$ by exploiting
correlations in $\psi(3770)\to D^0\xba D^0$. This provides crucial input
to the determination of $\gamma$ via the interference of $B^\pm\to K^\pm
D^0\to f$ with $B^\pm\to K^\pm \xba D^0\to f$. We discuss how the method
may be generalized to inclusive final states and illustrate with a toy
%AS1
model how such methods may offer one of the best means to determine
%AS1 end
$\gamma$ with $O(10^{8-9})$ B-mesons.

\newpage

\section{Introduction}\label{introduction}

The B factories at KEK and SLAC have made remarkable progress in many
areas of B-physics, in particular in the extraction of 
Cabibbo Kobayashi Maskawa
(CKM)~\cite{ckm} 
parameters crucial
to testing the standard model. Indeed, the determination of $\sin
2\beta$ via $B\to J/\psi K_S$ in such a way that there is no dependence on
theoretical assumptions 
promises to usher in a new era of precision tests of the CKM
paradigm\cite{belle_beta,babar_beta}.

The determination of the other two unitarity angles, $\alpha$ and
$\gamma$, without theoretical errors still presents a considerable
experimental challenge.  Even though CP violation itself may prove to be
easy to observe in channels sensitive to $\alpha$ and $\gamma$, the effect
tends to be modified by CP conserving effects such as strong phases.
To obtain a truly model independent determination of these angles, these
effects must be algebraically eliminated between several sets of
observables or measured in an independent experiment.

Final states containing $D^0, \xba D^0$ in decays of charged {\it and }
neutral B's provide methods for clean extraction of the angles of the
unitarity triangle (UT)\cite{pristine}. Direct CP violation in $B^{\pm}$
decays leads to a clean determination of $\gamma$. Time dependent CP
asymmetry measurements in decays such as $B^0$ or $\xba B^0 \to K^0 D^0
(\xba D^0)$ lead to a determination of $\delta \equiv \beta - \alpha + \pi
= 2 \beta + \gamma$\cite{branco,kayser,sanda,deltapaper} and also in fact
can give $\beta$\cite{kayser,deltapaper}. In these methods, for charged
and neutral B's, the crucial role is played by those final states of $D^0$
and $\xba D^0$ that are common to both. Indeed, in addition to the
exclusive modes of $D^0$, $\xba D^0$ that are common to them, as briefly
discussed in\cite{deltapaper} even multibody~\cite{ads2,scs2} and
inclusive decays can be used. The effectiveness of these $B \to K D$
methods for extracting angles of the UT can be vastly improved if charm
factory can provide some of the information for the relevant D decays.

The primary focus of this paper will be to consider the case of $\gamma$
from modes such as $B^\pm \to K^\pm (K^{*\pm}) D^0,\xba D^0$. Recall that
there are a number of different final states which are common to $D^0$ and
$\xba D^0$ decays that can be used here.  Such final states may be (1) CP
eigenstates (CPES) as first discussed in\cite{glw} (GLW) or CP
non-eigenstates (CPNES) as discussed in\cite{ads} (ADS).  The CPNES
may be (2) doubly Cabibbo suppressed (DCS) decays of $D^0$ (e.g. $K^+
\pi^-$) or (3) singly Cabibbo suppressed (SCS) states (e.g. $K^{*+} K^-$)
as recently considered in ~\cite{scs}. Amongst these, CP asymmetries
are expected to be large for case (2) and small for (1) and (3); on the
other hand, the relevant Br's tend to be larger for the latter two cases.

In the ADS method, the results from the interference through each $D^0$
channel depend on: (a) The CKM angle $\gamma$;  (b) The strong phase of
the $B$ decay; (c) The strong phase of the $D^0$ decay and (d) The decay
rate of $B^-\to K^- \xba D^0$.  The phase $\gamma$ of course is the
quantity that is of interest, and therefore, enough measurements, via
sufficient number of decay modes, must be made so that the dependence on
parameters (b) through (d) can be eliminated.

It has previously been suggested~\cite{ads,Sofferun} that charm factory data could
provide a useful 
additional input to the strong phase determination which is part of the 
ADS method. More generally,
there are a number of ways in which a charm factory~\cite{charmfactory}
can help in CKM-extractions.

\begin{enumerate}

\item Measurements of the Br's of some of the D decay modes
that enter the analysis; specifically determinations of the 
DCS D decays such as $K^+$ ($K^{*+}$) [$\pi^-, \rho^-, a_1^-$].

\item Improving the constraints on the $D^0 - \xba D^0$ mixing
parameters, $x_D$, $y_D$ can be very helpful.
\cite{ads2,Gronau:2001nr,ss1}.

\item More relevant to this paper is the use of a charm
factory to determine two parameters needed for being able to use 
(exclusive and) inclusive decays of $D^0, \xba D^0$, as briefly mentioned
in\cite{deltapaper}.  

\end{enumerate}

In particular this last application of charm factory data
offers the potential advantage of increasing the statistics available
in comparison to the original 
ADS method 
which 
only allows an analysis based on
exclusive states. 
Furthermore, it allows a way to integrate 3 or 4 body modes into the
analysis without making any assumption concerning the decay distribution
in phase space.

%
%
%------>describe subsequent sections here
%
%
%

In section~\ref{inclusive} we discuss a general formalism for strong
phases in inclusive states.  For exclusive states $f$ there is a single
strong phase difference $\ETA(f)$ between $D^0\to f$ and $\xba D^0\to f$.
We show that for an inclusive state $F$ we can introduce a net ``coherence
coefficient" $R_F$ together with a mean strong phase difference $\ETA(F)$ which
fully describes the relation between $D^0\to F$ and $\xba D^0\to F$.
In section~\ref{extracting} we discuss how these parameters can be
extracted from correlations at a $\psi(3770)$ factory. In
section~\ref{toy}, for illustration,
we define a toy model for the inclusive states $D^0\to
K^\pm+X$ and discuss the extraction of $\ETA$ and $R$. In
section~\ref{extracting_gamma} we discuss using this information to
extract $\gamma$ from $B^\pm\to K^\pm D^0$ with the subsequent decay of
$D^0$ to inclusive states along with the $\psi(3770)$ factory determination
of $\ETA$ and $R$ and we give some numerical results obtained
with the toy model. A brief summary appears in section VI.

\section{Inclusive States}\label{inclusive}

In the following discussion we will define an exclusive 
decay of the $D^0$ to be any decay which is governed by a single quantum
mechanical amplitude. Thus, a decay such as $D^0\to K^-\pi^+$ would be an
exclusive state while for a three body decay such as $D^0 \to K^-\pi^+\pi^0$,
each point on the Dalitz plot would be considered a distinct exclusive
state. 

Conversely, an inclusive state is any state which is a set of
exclusive states. For instance $D^0\to K^-\pi^+\pi^0$ integrated over all
or part of the Dalitz plot would be an inclusive state as sets of states
with different particle content such as $D^0 \to K^- + n\pi$. Inclusive
states defined in this way may either be composed  of a collection
of discrete states (eg. $\{ K^-+\pi^+$, $K^-+\rho^+\}$), of states which
are a continuum (eg $K^-\pi^+\pi^0$ integrated over the Dalitz plot)   
or of states which are a combination of both (eg. $K^-+n\pi$). In the
discussion below we will treat an inclusive state as being composed of
a discrete set of exclusive states although the generalization to
continuous sets of states is straightforward. Thus if $F$ is an inclusive
final state we will write:

\begin{eqnarray}
F&=\{ f_i \}
\end{eqnarray}

\noindent
where $f_i$ are the exclusive states which make up $F$.

Primarily we are interested in final states which are common to $D^0$ and
$\xba D^0$ decay. Two categories of such final states of particular
interest are DCS states such as $K^+\pi^-$ and CP eigenstates such as
$K_S\pi^0$. For each $f_i\in F$, let us denote:

\begin{eqnarray}
A(f_i)&=&{\cal M}(D^0\to f_i)\nonumber\\
\xba A(f_i)&=&{\cal M}(\xba D^0\to f_i)\nonumber\\
\ETA(f_i)&=&\arg(A^*(f_i)\xba A(f_i))
\end{eqnarray}

\noindent so $\ETA(f_i)$ is the strong phase difference between $D^0$ and
$\xba D^0$ decay to $f_i$.

Because both the channels $B^-\to K^- D^0$ and $B^-\to K^- \xba D^0$
can contribute to the overall process $B^-\to K^- F$ for the inclusive
state $F$, we can regard the object that decays into $F$ do be a quantum
mechanical mixture of $D^0$ and $\xba D^0$.

Denoting this state by $|I>$, then

\begin{eqnarray}
|I\rangle=a |D^0\rangle+be^{i\lambda}|\xba D^0\rangle
\end{eqnarray}

\nonumber
where $a$ and $b$ are real.

We can thus expand the decay rate for the mixed state $|I>$ to $F$ 
as:

\begin{eqnarray}
\Gamma(I\to F)
&=&
\sum_i
\bigg\{
a^2 |A(f_i)|^2
+
b^2 |\xba A(f_i)|^2\nonumber\\
&&+
2ab|A(f_i)|~|\xba A(f_i)|\cos(\ETA(f_i)+\lambda)\bigg\}
\nonumber\\
&=&
a^2 A(F)^2
+
b^2 \xba A(F)^2\nonumber\\
&&+
2R_FabA(F)\xba A(F)\cos(\ETA(F)+\lambda)
\label{sum_amps}
\end{eqnarray}

\noindent
where

\begin{eqnarray}
A(F)^2&=&\sum |A(f_i)|^2 \nonumber\\
\xba A(F)^2&=&\sum |\xba A(f_i)|^2 \nonumber\\
R_Fe^{i\ETA(F)}&=&{\sum |A(f_i)||\xba A(f_i)|e^{i\ETA(f_i)}\over A(F)\xba
A(F)}
\end{eqnarray}

\noindent
The key point to note is that $R_F$ and $\ETA(F)$ are independent of $a$,
$b$ and $\lambda$. The decay rate of $I$ thus depends only on four
parameters of $F$, namely $A(F)$, $\xba A(F)$, $R_F$ and $\ETA_F$,
regardless of how many states make up $F$. We can think of $A(F)$ and
$\xba A(F)$ as the average amplitudes of $D$ and $\xba D$ decay to $F$
while $\ETA(F)$ is the average strong phase difference for $F$ and $R_F$
is a measure of the coherence of $F$. Note that $0\leq R_F\leq 1$.

In the case where $F=\{f_1\}$ consists of a single quantum state, then
$R_F=1$. In this case the decay rate of $I$ is only determined by three
parameters: the amplitudes of $D^0$ and $\xba D^0$ decay and the strong
phase difference, $\ETA(F)=\ETA(f_1)$. If $f_1$ is a CP eigenstate then,
assuming that $D^0$ decay is CP conserving\cite{nocp}, 
$A(F)=\xba A(F)$, $R_F=1$ and
$\ETA(F)=0$ or $\pi$ depending on whether $f_1$ is CP=$+1$ or CP=$-1$.

More generally if $F$ is a set of states such that $CP:F\to F$ then it
follows that $A(F)=\xba A(F)$ and $\ETA(F)=0$ or $\pi$ (depending on
whether $F$ is predominantly CP=$+1$ or CP=$-1$) but $R_F$ depends on the
makeup of $F$ and indicates the purity of $F$. Thus, if $F$ is made up of
CP eigenstates with the same CP=$\pm 1$ eigenvalue, $R_F=1$ and $F$
behaves as a single CP=$\pm 1$ eigenstate. In \cite{ads2,scs2}
the extraction of $\gamma$ using the detailed analysis of 3 and 4 body
states of this form is considered via an analysis of the amplitude
structure in phase space.

Some examples of 
inclusive states to which we can apply this approach are:

\begin{enumerate}

\item $F=\{K^-\pi^+\}$: This is a single CP non eigenstate (CPNES) so
therefore $R_F=1$ while $A(F)$, $\xba A(F)$ and $\ETA(F)$ need to be
determined experimentally.  Current measurements~\cite{PDB} of the
branching ratios give $Br(D^0 \to K^-\pi^+)= 3.80\% $ and 
$Br(D^0 \to
K^+ \pi^-) \approx 1.5\times 10^{-4} $ 
thus $\xba A(F)/A(F) \approx 0.05$ in this case.

\item $F=\{K^-\pi^+\pi^0\}$ This is an inclusive CPNES if one integrates
over the Dalitz plot; therefore $A(F)$, $\xba A(F)$, $\ETA(F)$ and $R_F$
need to be experimentally determined. It is useful to break
down the Dalitz plot into a number of sub-regions $F=F_1\cup F_2\cup \dots
\cup F_n$ each of which will be characterized by $A(F_i)$, $\xba A(F_i)$,
$\ETA(F_i)$ and $R_{F_i}$

\item $F=\{K^-+X\}$: This case is an even more inclusive CPNES than the
previous one. Again it may be useful to decompose F into a number of
subsets determined either by the composition of $X$ (eg. number of pions)
or by the energy of the $K^-$.

\item $F=\{ K_s \pi^0\}$: This is a single CP eigenstate (CPES)  with
CP=$-1$ therefore $A(F)=\xba A(F)$, $R_F=1$ and $\ETA(F)=\pi$.

\item $F=K_s+X$: This is a CP invariant inclusive state (CPIIS) so
$A(F)=\xba A(F)$ and $\ETA_F=0$ or $\pi$. Which value of $\ETA_F$ applies
and the value of $R_F$ need to be determined experimentally. Again, it may
be useful to decompose this set of states according to the composition of
$X$ or the energy of the $K_S$. Note that changing the $K_S$ to a $K_L$
changes $\ETA(F)\to \pi-\ETA(F)$ but keeps the other parameters unchanged.

\item $F=K_S\pi^+\pi^-$: Again this is a CPIIS which can be decomposed on
the basis of the energy of the $K_S$.  

\end{enumerate}

\section{Extracting Phases and Coherence from
$\psi(3770)$}\label{extracting}

The ability to use inclusive and exclusive decays of $D^0$ in order to
obtain CP violation phases will be enhanced if a separate determination of
$A(F)$, $\xba A(F)$, $R_F$ and $\ETA(F)$ can be made.

We will assume that $A(F)$ and $\xba A(F)$ can be determined from the
$D^0$ branching ratios.  A determination of $\ETA(F)$ and $R_F$ may be
made at a $\psi(3770)$ charm factory.

This follows from the fact that  $\psi(3770)$ is a spin-1 
state and therefore the decay $\psi\to D^0\xba D^0$ is an antisymmetric
wave function:

\begin{eqnarray}
(|D^0\rangle |\xba D^0\rangle
-
|\xba D^0\rangle | D^0\rangle
)/\sqrt{2}
\label{bose}
\end{eqnarray} 

\noindent
This entangled state gives us access to strong phase information for
final states in a number of different ways. Here we will discuss the case
where we assume $D^0\xba D^0$ oscillation is small.
%
% , in the Appendix we
% will generalize to the case where oscillation is present. 
%

We can take advantage of this entanglement by observing various
correlations between the decay of the two mesons which arise from
$\psi(3770)$ decay. For a given inclusive state $F$, it is useful to
distinguish between 4 different kinds of correlations

\begin{enumerate}
\item The correlation of $F$ with another inclusive state $G$ by measuring
the branching ratio $\psi(3770)\to [F] [G]$

\item The correlation of $F$ with its charge conjugate by measuring the
branching ratio $\psi(3770)\to [F] [\xba F]$
as previously discussed in the case of exclusive states in~\cite{Sofferun,ss1}.

\item The correlation of $F$ with a CP eigenstate by measuring the
branching ratio $\psi(3770)\to [F] [CP\ eigenstate]$

\item The correlation of $F$ with itself by measuring the branching ratio
$\psi(3770)\to [F] [F]$

\end{enumerate}

%
%
%%%%%%%%%%%%%%%%%%%%%%%%%%%%%%%%%%%%%%%%%%%%%%%%%%%%%%%%%%%%%%%
%
%

%
%
%
%
%
%%%%%%%%%%%%%%%%%%%%%%%%%%%%%%%%%%%>>>>>>>start rewrite section
%
%
%
%
%

Let us consider first the most general case where the $\psi(3770)$ 
decays overall to the final state $FG$ where $F$ and $G$ are inclusive
final states. 

The decay rate to this final state is thus

\begin{eqnarray}
\Gamma(FG)&=&
\Gamma_0
\sum_{ij}
\left | 
A(f_i) \xba A(g_j)
-
\xba A(f_i)  A(g_j)
\right |^2
\nonumber\\
&=&
\Gamma_0
\sum_{ij}
\bigg [
|A(f_i)|^2
|\xba A(g_j)|^2
+
|\xba A(f_i)|^2
|A(g_j)|^2
\nonumber\\
&&
-
2 
|A(f_i)
A(g_j)
\xba A(f_i)
\xba A(g_j)|
\cdot
\nonumber\\
&&\cdot
\cos(\ETA(f_i)-\ETA(g_j))
\bigg ]
\end{eqnarray}

\noindent where $\Gamma_0=\Gamma(\psi(3770)\to D^0\xba D^0$. Summing this
over $i$ and $j$ we thus obtain an expression in terms of the exclusive
quantities:

\begin{eqnarray}
\Gamma(FG)
&=&
\Gamma_0
\bigg[
A_F^2\xba A_G^2
+
\xba A_F^2  A_G^2
\nonumber\\
&&-2 R_F
R_G
A_F 
\xba A_F 
A_G
\xba A_G
\nonumber\\
&&
\cos(\ETA(F) - \ETA(G))
\bigg]
\label{FG}
\end{eqnarray}

A special case of the above is where $G=\xba F$ in which case this
expression reduces to:

\begin{eqnarray}
\Gamma(F\xba F)&=&
\Gamma_0
\bigg[
A_F^4
+
\xba A_F^4
-2 R_F^2
A_F^2 
\xba A_F^2 
\cos(2\ETA(F))
\bigg]
\label{FFbar}
\end{eqnarray}

If, on the other hand, $G$ is a CP eigenstate, or indeed a set of CP
eigenstates with the common eigenvalue $\lambda_{CP}=\pm 1$, then
$\ETA_G=0$ or $\pi$ respectively and $R_G=1$ while $A(G)=\xba A(G)$. 
In this case then,

\begin{eqnarray}
\Gamma(FG)
&=&
\Gamma_0A_G^2
\bigg[
A_F^2
+
\xba A_F^2  
\nonumber\\
&&-2 \lambda_{CP} R_F
A_F 
\xba A_F 
\cos(\ETA(F))
\bigg]
\label{FCP}
\end{eqnarray}

Finally, in the special case where $F=G$ then

\begin{eqnarray}
\Gamma(FF)
&=&
\Gamma_0
A_F^2\xba A_F^2(1-R_F^2)
\label{FF}
\end{eqnarray}

\noindent
Note that this expression gives 0 if $F$ is an exclusive state as expected
by Bose symmetry~\cite{Gronau:2001nr}.
In each of these cases, one can specialize to the case where $F$ (or $G$)
is an exclusive case where we then have $R_F=1$ (or $R_G=1$).

From the above relations, it is clear that if we have a number of
inclusive or exclusive states we can solve for the various values of
$\ETA(F)$ and $R_F$.  To make this clearer, for a given set of inclusive
final states ${\cal F}=\{F_1,\dots F_n\}$ let us define $n_o$ to be the
number of observables, $n_p$ the number of parameters and $\delta
n=n_o-n_p$. Thus, $\delta n \geq 0$ is a necessary condition for the
determination of the parameters.

If ${\cal F}$ contains $k$ inclusive states and $\ell$ exclusive states,
$n=k+\ell$, then the total number of free parameters is $n_p=2k+\ell$,
i.e. $\ETA$ for each inclusive and exclusive state and $R$ for each
inclusive state.

For each pair of distinct members (exclusive or inclusive)  of ${\cal F}$
there are two distinct
observables, $Br(FG)=Br(\xba F \xba G)$ and $Br(F\xba G)=Br(\xba F G)$.
This gives a total contribution to $n_o$ of $n(n-1)$.

For each inclusive member of ${\cal F}$, there are the two observables
$Br(FF)=Br(\xba F \xba F)$ and $Br(F \xba F)$ 
giving a contribution to $n_o$ of
$2k$ while for each exclusive state $Br(FF)\equiv 0$, hence the
contribution is $\ell$ from correlations of the form $F\xba F$. 

Finally, for each member of ${\cal F}$ one can observe the correlation
with a set of CP eigenstates giving an additional contribution of $n$ to 
$n_o$. Thus, 

\begin{eqnarray}
n_p &=& 2k+\ell  =k+n
\nonumber\\
n_o &=& k+n+n^2
\nonumber\\
\delta n &=& n^2
\label{numobs}
\end{eqnarray}

\noindent
where $n_o$ is the sum of $n(n-1)$ correlations of the form $FG$ and
$F\xba G$; $n$ of the form $F\xba F$; $n$ of the form $F$ + $CPES$ and $k$
of the form $FF$ (for inclusive states only).

Clearly, using one state ``in isolation'' Eqn.~(\ref{FF}) and
Eqn.~(\ref{FFbar})  give us just enough information to determine $R$ and
$\ETA$ for an inclusive state while Eqn.~(\ref{FFbar})  will give us
information to determine $\ETA$ for an exclusive state. Observing the
cross correlations with other states and CP eigenstates thus gives us a
degree of over determination $\delta n=n^2$.

Of course if the branching ratio to some of the final states is small, 
the statistical error on some of these correlations may be large. To
consider how well this program may be carried out, we will construct a toy
model of certain $D$ decays.

%
%
%%%%%%%%%%%%%%%%%%%%%%%%%%%%%%%%%%%%%%%%%%%%%%%%%%%%%%%%%%%%%%%
%
%

\section{Toy Models}\label{toy}

The effectiveness of various methods in determining $\delta$ depends to
some extent on the properties of various $D$ decays. Since these decays
have not been fully characterized, particularly in the DCS modes, we will
use a ``toy model'' in which we will endeavor to capture the known
properties of these decays. This should allow us to obtain a general idea
of the values of $R$ and $\ETA$ which will be obtained for various classes
of inclusive states.  In particular we will construct a toy model for 
$K^\pm+X$ final states.
We will attempt to model at least part of the rate in each of these
channels as the sum of exclusive states. Each exclusive final state may
either be produced through a continuum process, in which case we will
assume that the amplitude is constant over phase space, or it may be
produced through resonance channels which we will describe by a
Breit-Wigner distribution.

\subsection{Model for $D^0\to K^-+X$ and $K^++X$}

In this section we consider a model for $\F$, the set of $K^-+X$ where $X$
contains at most one $\pi^0$.  We will model such decays by considering
decays of the form $D^0\to K^-\pi^+$; $D^0\to K^-\pi^+\pi^0$; $D^0\to
K^-\pi^-\pi^+\pi^+$ and $D^0\to K^-\pi^-\pi^+\pi^+\pi^0$.

In Table~\ref{tab1} we give the decomposition of these modes into
resonance channels which we consider and their branching ratios
from~\cite{PDB}. Note that the sum of all these branching ratios is 29\%
out of the total 58\% for all $K^-+X$ states. In~\cite{D3k_model} the
experimental data for $D^0\to K^-\pi^+\pi^0$ was fit to a model with
$K^{0*}$, $\rho^+$ and $K^{+*}$ channels and a 3-body continuum. We will
use this model to 
describe these 3-body decays.

For each of the quasi-two body modes, we will assume that the resonances
(ie. $\rho$, $a_1$, $K^*$) is modeled by a Breit-Wigner amplitude while
the continuum 3 and 4-body states we will assume have a constant amplitude
over phase space. We will assume that the strong phase difference between
the different channels that lead to $K+3\pi$ and $K+4\pi$ final states are
$0$ although we find that the results considered below do not depend
greatly if arbitrary phase differences are used.

For the corresponding DCS decays $\xba D^0\to K^- + X$ we
apply SU(3) to the Cabibbo-allowed 
decays, in particular we exchange $d\leftrightarrow
s$ in the final state. We rescale the amplitudes so that the total DCS
rates match the results in~\cite{PDB}.

\begin{figure}
\epsfxsize 3.0 in
\mbox{\epsfbox{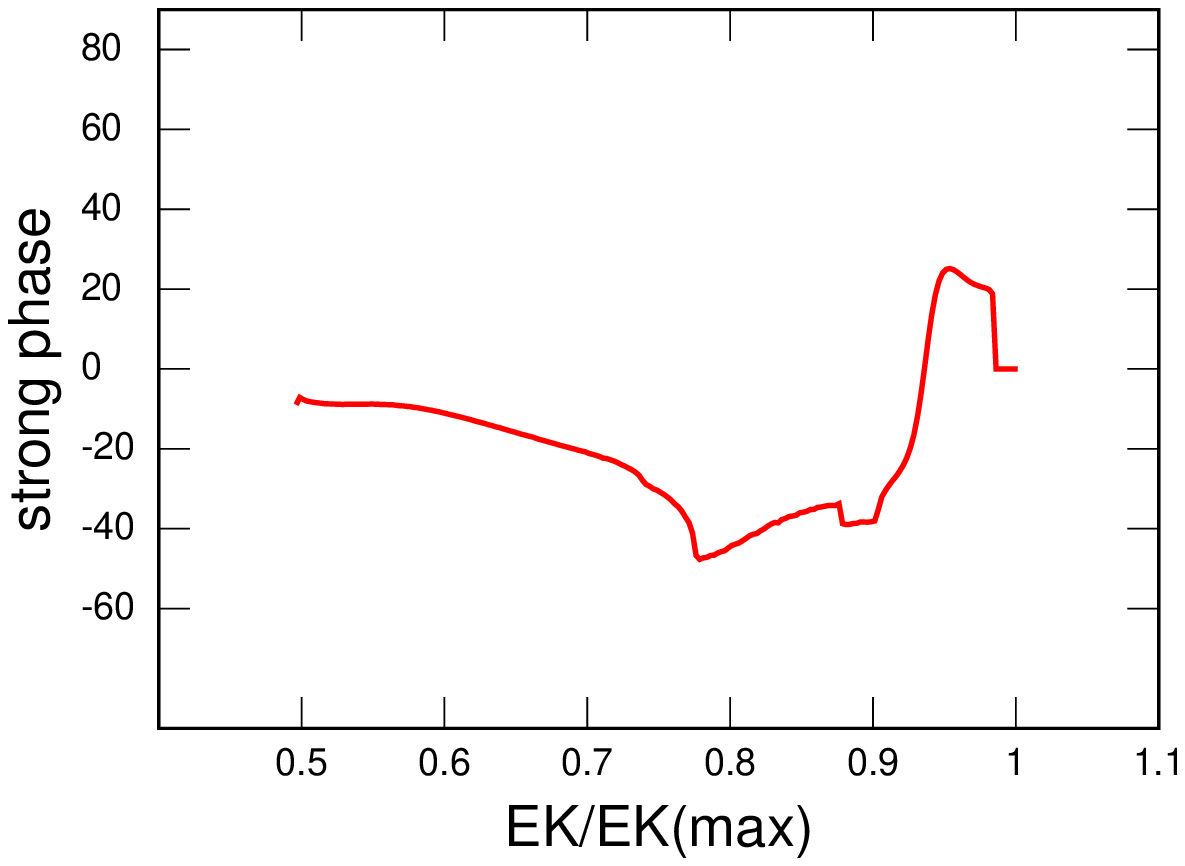}}
\caption{
$\ETA$ is shown as a function of the energy fraction
of the $K^-$
for our model of $\F$.
}\label{eta_graph}
\end{figure}

\begin{figure}
\epsfxsize 3.0 in
\mbox{\epsfbox{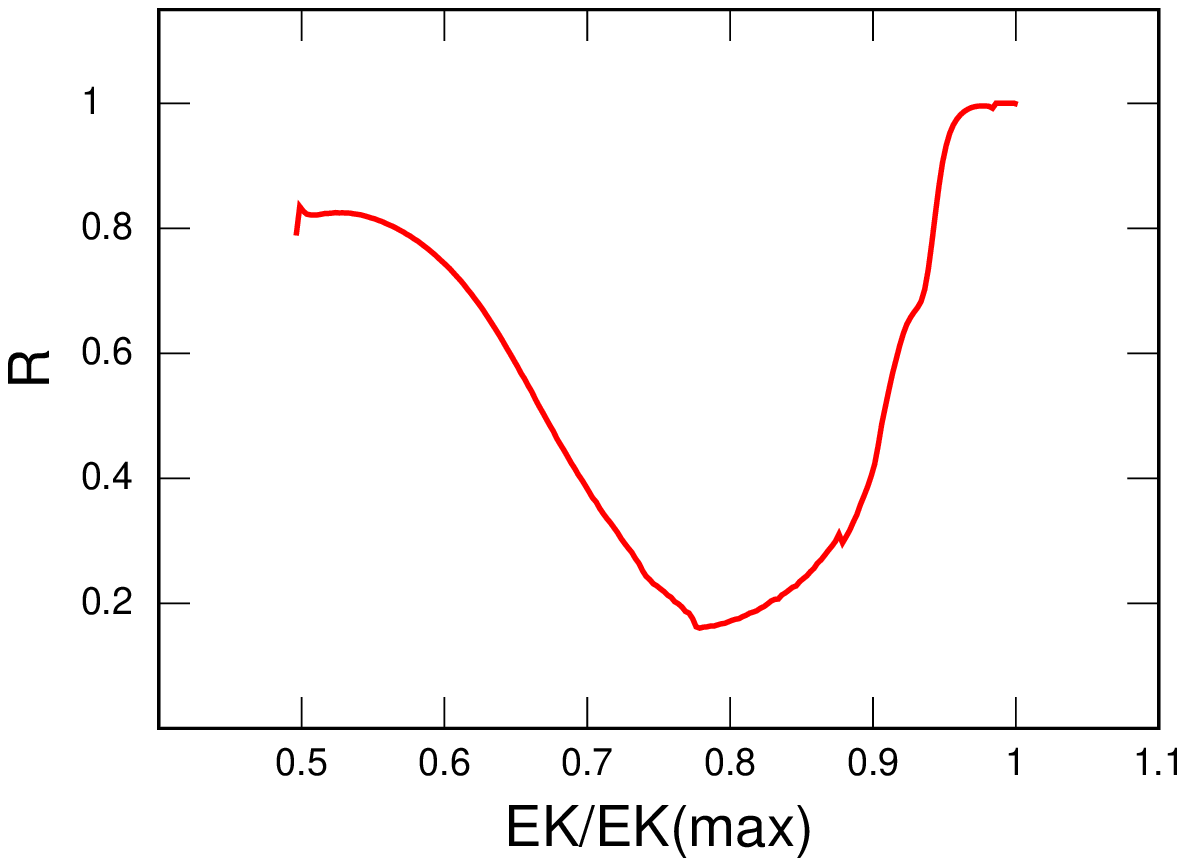}}
\caption{
$R$ is shown as a function of the energy fraction
of the $K^-$
for our model of $\F$.
}\label{R_graph}
\end{figure}

Using this model, we find that for $\F$ in total, $R_\F=0.51$ and
$\ETA_\F=-11^\circ$. In Fig.~\ref{eta_graph}, we show $\ETA$ as a
function
of the energy fraction $E_K/E_{max}$ of the $K^-$ where
$E_{max}=(m_D^2+m_K^2-m_\pi)/(2m_D)$.
Likewise in Fig.~\ref{R_graph} we show $R$ as a function of the energy
fraction. 

From Fig.~\ref{R_graph} we see that intermediate values of $E_K$ have
smaller values of $R$ than low energy and high energy $K$'s. We therefore
consider dividing $\F$ into three subsets $\F=\F_1\cup\F_2\cup\F_3$ where
for $\F_1$, $E_K/E_{max}\leq 0.65$ for $\F_2$ $0.65\leq E_K/E_{max}\leq
0.9$ and for $\F_3$ $0.9\leq E_K/E_{max}$.

Using the model, we find that 
for $\F_1$, $\ETA_{\F_1}=-34^\circ$
and $R=0.74$ while $Br(\F_1)/Br(\F)=20.5\%$;
for $\F_2$, $\ETA_{\F_2}=-86^\circ$
and $R=0.29$ while $Br(\F_1)/Br(\F)=42.9\%$
and
for $\F_3$, $\ETA_{\F_3}=30^\circ$
and $R=0.91$ while $Br(\F_1)/Br(\F)=36.4\%$.

Likewise we can consider subsets of $\F$ determined by particle content.
For example if we define $\F_{3bdy}$ to be final states of the form
$K^-\pi^0\pi^-$ and $\F_{4bdy}$ to be final states of the form
$K^-\pi^-\pi^+\pi^+$ then $\ETA_{\F_{3bdy}}=-2.1^\circ$; $R_{{3bdy}}=0.60$
while $\ETA_{{4bdy}}=-81^\circ$; $R_{{4bdy}}=0.13$.

The degree of CP violation in $B^-\to K^- [D^0\to F]$ 
is proportional to $R_F$ so a larger value of $R$, in general, indicates
greater utility in terms of extracting $\gamma$. The strong phase $\ETA$
will be combined with the strong phase difference between 
$B^-\to K^- D^0$
and 
$B^-\to K^- \xba D^0$.

%
%
% \subsection{Model for $D^0\to K^++X$}
%
%
% \subsection{Model for $D^0\to$ CP eigenstates}
%
%
% \subsection{Model for $D^0\to K_{L,S}+X$}
%
%

\subsection{Phase and Coherence Determination for Toy Model}

Let us consider now the determination of $\ETA$ and $R$ at a $\psi$
factory for toy model for $K+X$.  In the following, we will use the above
decomposition of the events into $\F_1$, $\F_2$ and $\F_3$ according to
the $K^\pm$ energy.

The CP eigenstates which we will consider will consist of 
($CP=-1$) 2 body final states which do not contain a $K_L$. The branching
ratios to such final states is $\sim 5\%$. Clearly some improvement could
be obtained if more general final states were considered.

%>>>>>>>>>>>>>>>>>>>>>>>>>>>>>>>here

To determine the 6 parameters $R_{1-3}$ and $\ETA_{1-3}$ we thus have 
the modes displayed in Table~(\ref{mod_parm})
which, according to Eqn.~(\ref{numobs}) gives us $n_o=15$ independent
observables for $n_p=6$ parameters giving $\delta n=n^2$.

\begin{figure}
\epsfxsize 3.0 in
\mbox{\epsfbox{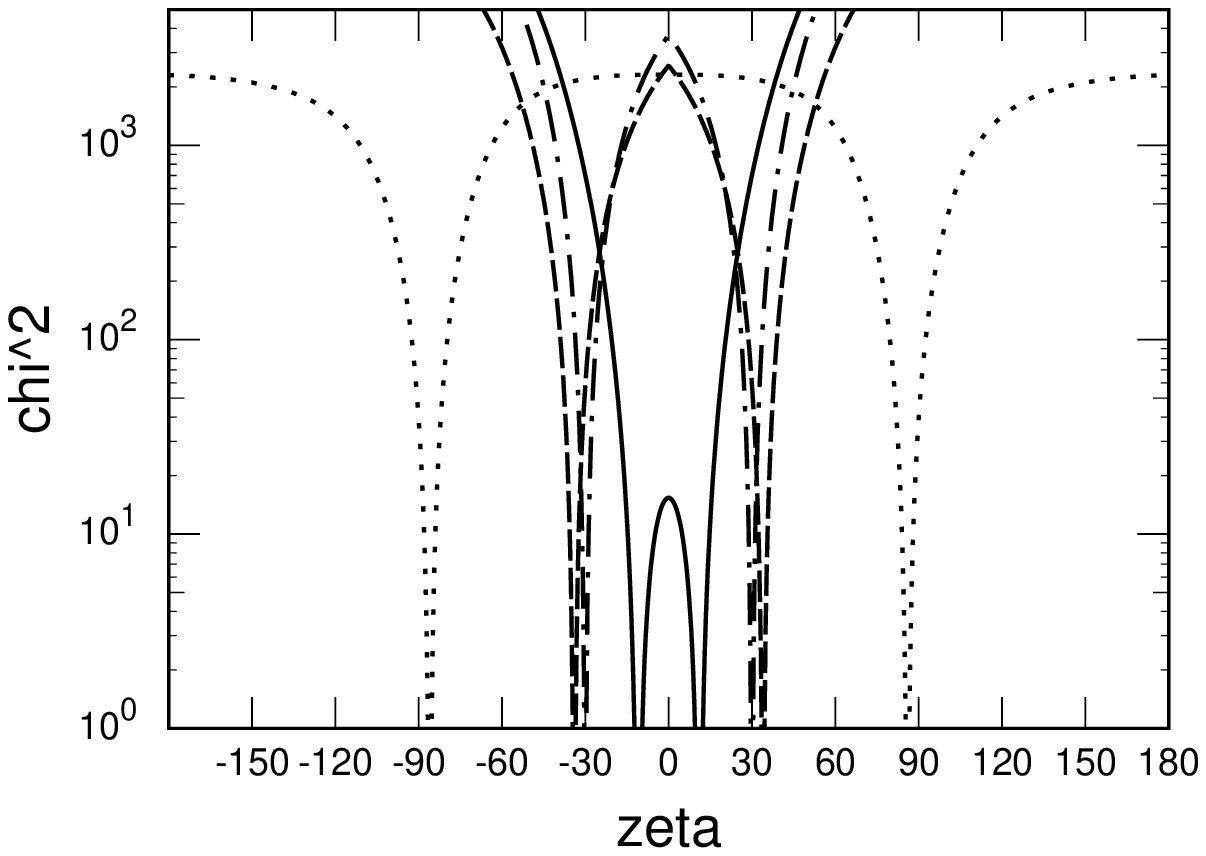}}
\caption{
$\chi^2$ is plotted as a function of $\ETA$ 
using correlations at a $\psi(3770)$ factory for the inclusive 
modes $\F_1$, $\F_2$ and $\F_3$ defined in the text. 
Using the 15 correlations of the form $\F_i\F_j$, $\F_i\bar F_j$ and
$\F_i$ with CPES-, the dashed curve gives $\chi^2$ as a function of 
$\ETA(\F_1)$, the dotted curve gives $\chi^2$ as a function of
$\ETA(\F_2)$ and the dash-dotted curve gives $\chi^2$ as a function of
$\ETA(\F_3)$. The solid curve gives $\chi^2$ as a function of $\ETA(\F)$
using only the correlations of $\F\F$; $\F\bar\F$ and $\F$ with CPES-.
}\label{inclusive_eta}
\end{figure}

In order to estimate how  much data is required to determine the
parameters $\ETA_i$ and $R_i$, we suppose that there are $\hat
N_{DD}=10^7$ events~\cite{charmfactory} of the form 
$\psi(3770)\to D^0 \bar 
D^0$. In
Fig.~(\ref{inclusive_eta})  the solid curve indicates the result for $\F$
taken as a whole so that the data which is used is the correlation of $\F$
with $\F$, $\F$ with $\xba \F$ and $\F$ with CPES-.
The other three curves indicate the results for $\F_1$ (dashed), $\F_2$
(dotted) and $\F_3$ (dash-dotted).

The resulting curves are invariant under the transformation $\ETA_i\to
-\ETA_i$ and $\ETA_i\to \pi+\ETA_i$ since those transformations clearly
leave the correlations unchanged.  The angles $\ETA_i$ can only therefore
be determined up to a 4 fold ambiguity.

From these curves it is clear that modulo the ambiguity, the angles are
determined to $O(2^\circ)$ with $\hat N_{DD}=10^7$ events.

\begin{figure}
\epsfxsize 3.0 in
\mbox{\epsfbox{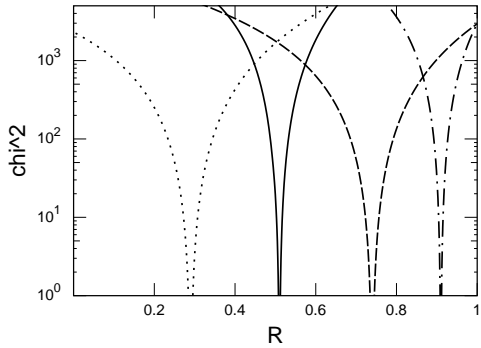}}
\caption{
$\chi^2$ is plotted as a function of $R$
using correlations at a $\psi(3770)$ factory 
for the same final states 
as in Fig.~(\ref{inclusive_R}).
}\label{inclusive_R}
\end{figure}

In Fig.~(\ref{inclusive_R}) a similar plot of the minimum value of
$\chi^2$ is given as a function of $R_i$ for $\F_i$ given $N_{DD}=10^7$.  
In this case $R_i$ can be determined to $\sim 2\%$ with $N_{DD}=10^7$
while with $N_{DD}=10^6$ the 3-sigma bound on R is about $10\%$. We will
find that applying this determination of $\ETA$ and $R$ is more than
adequate for the determination of $\gamma$ at the $B$ factory.

%%%%
%
%---------------------------------------------------------------
%
%%%%

\section{Extracting $\gamma$ from Direct CP
Violation}\label{extracting_gamma}

Let us now turn our attention to the case of $B^\pm\to D^0 K^\pm$. In this
case the two contributing amplitudes are $b\to c\xba u s$ and $b\to u\xba
c s$ and so the sensitivity is only to the single CKM angle $\gamma$ and
indeed this process has been discussed extensively~\cite{glw,ads,scs} as
a means to measure this angle.  The crucial factor in an effective
determination of $\gamma$ is the final state chosen for the $D^0$ decay.

The method proposed in~\cite{glw}  (GLW method)
requires 
the following
measurements, two of which (eqn.~(\ref{glw_hard1}) and 
eqn.~(\ref{glw_hard})) are the same:

\begin{eqnarray}
Br(B^+\to K^+ [D^0\to {\rm FES}-])\label{glw_easy1}\\
Br(B^-\to K^- [D^0\to {\rm FES}+])\label{glw_easy}\\
Br(B^+\to K^+ [D^0\to {\rm FES}+])\label{glw_hard1}\\
Br(B^-\to K^- [D^0\to {\rm FES}-])\label{glw_hard}\\
Br(B^-\to K^- [D^0\to {\rm CPES}])\label{glw_cpesa}\\
Br(B^+\to K^+ [D^0\to {\rm CPES}])\label{glw_cpesb}
\end{eqnarray}

\noindent where $FES\pm$ is a flavor eigenstate of charm=$\pm 1$ and $D^0$
generically means a mixture of $D^0\xba D^0$.  One expects an O(10\%) CP
violating difference between (\ref{glw_cpesa}) and (\ref{glw_cpesb})  
from this data; it is possible to reconstruct $\gamma$ up to an 8-fold
ambiguity.

There is, however, a practical problem with the observation of FES states.
The only states which are pure FES's are semi-leptonic decays such as
$D^0\to \ell^+\nu_\ell K^-$. In the case of the reaction
Eqn.~(\ref{glw_hard}) there is the potential backgrounds from semileptonic
decay of the parent $B$~\cite{ads}. There is perhaps some prospect of
overcoming this in the case of the $D^{*0}$ analog~\cite{sokoloff} but
this also may not be easy.

Cabibbo allowed hadronic final states cannot be used at all because in all 
such cases
reaction~(\ref{glw_easy}) followed by a DCS decay to the same final state
will quantum mechanically interfere. In fact such interference is O(100\%)
and thus provides another route to the determination of $\gamma$.

In particular, if $f$ and $g$ are exclusive final states
(one of them may be a CP eigenstate) 
then $\gamma$ may
be determined~\cite{ads} 
(ADS method)
up to a 8-16 fold ambiguity from:

\begin{eqnarray}
Br(B^-\to K^- [D^0\to f]      \label{adsf}\\
Br(B^+\to K^+ [D^0\to \xba f] \label{adsfx}\\
Br(B^-\to K^- [D^0\to g]       \label{adsg}\\
Br(B^+\to K^+ [D^0\to \xba g]  \label{adsgx}
\end{eqnarray}

\nonumber where we assume that the branching ratios of $D$ and $\xba D$ to
$f$ and $g$ are known and that $Br(B^-\to D^0 K^-)$ is known but we must
fit for the branching ratio $Br(B^-\to \xba D^0 K^-)$ which seems
difficult to measure experimentally. For the method to be effective, $f$
should be chosen such that $D^0\to f$ is a DCS transition in which case
the CP violating difference between reaction (\ref{adsf}) and
(\ref{adsfx}) is expected to be O(100\%).

To obtain a good determination of $\gamma$ it
is best to add additional  modes which will therefore over determine
$\gamma$ and reduce the ambiguity to 4-fold (i.e. a unique determination
of $\sin^2\gamma$).

Let us now generalize this method by considering inclusive final states.
As we discussed above, each inclusive set, $F$, carries with it an
additional parameter $R_F$ compared to an exclusive final state.  With
such final states, one can never hope to determine $\gamma$ since the
reactions

\begin{eqnarray}
Br(B^-\to K^- [D^0\to F]      \label{adsF}\\
Br(B^+\to K^+ [D^0\to \xba F] \label{adsFx}
\end{eqnarray}

\noindent provide two observables but introduce the two additional degrees
of freedom $\{\ETA_F,R_F\}$. One must therefore have additional
information bearing on these parameters. 
Clearly then, 
data from a $\psi(3770)$ charm
factory can directly determine $R_F$ and $\ETA_F$ for each inclusive
final state which can provide the additional information required.

%-----------------------------------------------------------

\subsection{Inclusive ADS}

Following~\cite{ads}
let us introduce the following notation for the
various branching ratios:

\begin{eqnarray}
a=Br(B^-\to k^- D^0) &~& \xba a=Br(B^+\to k^+ \xba D^0)
\nonumber\\
b=Br(B^-\to k^- \xba D^0) &~& \xba b(k)=Br(B^+\to k^+ D^0)
\nonumber\\
c(F)=Br(D^0\to F) &~&  \xba c( F)=Br( \xba D^0\to F) 
\nonumber\\
c(\xba F)=Br(D^0\to \xba F) &~&  \xba c( \xba F)=Br( \xba D^0\to \xba F)
\nonumber
\end{eqnarray}
\begin{eqnarray}
d(k,F)&=&Br(B^-\to k^- [D^0\to F]) 
\nonumber\\
\xba d(k,\xba F)&=&Br(B^+\to k^+ [D^0 \to \xba F])
\end{eqnarray}

\noindent   
Here $k^\pm$ represents either $K^\pm$ or $K^{*\pm}$ (or indeed one may
consider any other kaonic resonance or system of strangeness=$-1$ and well
defined CP).

In the standard model, it is expected that $a(k)=\xba a(k)$, $b(k)=\xba
b(k)$ and $\xba c(X)= c(\xba X)$ all of which we will assume 
from here on~\cite{nocp}.
The value of the quantities $d$, $\xba d$ may be expressed in terms of
$a$, $b$ and $c$ as:

\begin{eqnarray}
d(k,F)&=& a(k)c(F)+b(k)c(\xba F)
\nonumber\\
&&+2R_F\sqrt{a(k)b(k)c(F)c(\xba F)}
\cdot
\nonumber\\
&&\cdot
\cos(\zeta_k+\ETA_F+\gamma)
\nonumber\\
\xba d(k,\xba F)&=& a(k)c(F)+b(k)c(\xba F)
\nonumber\\
&&+2R_F\sqrt{a(k)b(k)c(F)c(\xba F)}
\cdot
\nonumber\\
&&\cdot
\cos(\zeta_k+\ETA_F-\gamma)
\label{eqnd}
\end{eqnarray}

\noindent where $\zeta_k$ is the strong phase difference between $B^-\to
k^- D^0$ and $B^-\to k^- \xba D^0$; $\ETA_F$ is the strong phase
difference between $D\to F$ and $D\to \xba F$ and $\gamma$ is the CP
violating weak phase difference between $B^-\to k^- D^0$ and $B^-\to k^-
\xba D^0$.

%
%
%---------------------------------begin new stuff
%
%

To illustrate the procedure for finding $\gamma$, let us first consider
the use of exclusive modes via the ADS method and then supplementing it
with information obtained from a $\psi(3770)$ factory. The exclusive modes
we sill consider are 

\begin{enumerate}
\item $D^0\to K^+\pi^-$
\item $D^0\to K^{*+}\pi^-$
\item $D^0\to$  CP=$-1$ eigenstates (CPES-).
\end{enumerate}

The branching ratios for $\xba D^0\to K^+\pi^-$ is 3.80 \% while the
branching ratio for $D^0\to K^{+}\pi^-$ is $1.48\times 10^{-4}$. The
branching ratio for $\xba D^0\to K^{*+}\pi^-$ is $6.0\%$ while the
corresponding branching ratio for the $D^0$ decay is unknown. For our
calculation we will assume that it is the latter multiplied by the double
Cabibbo suppression factor $\sin^2\theta_c$. The branching ratio to
$CP=-1$ states we will take to be about $5\%$. Of course the strong phases
and $\gamma$ are totally unknown but for the purposes of illustration, we
will take as the real value of $\gamma=60^\circ$, with
$\ETA(K^+\pi^-)=120^\circ$; $\ETA(K^{*+}\pi^-)=60^\circ$ and
$\ETA_B=-50^\circ$.

\begin{figure}
\epsfxsize 3.0 in
\mbox{\epsfbox{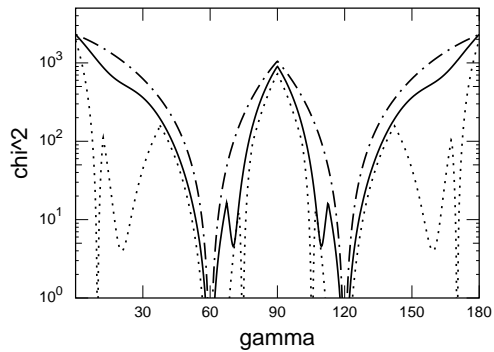}}
\caption{
$\chi^2$ is plotted as a function of $\gamma$ for input generated assuming
$\gamma=60^\circ$ and $\zeta_k=-50^\circ$ 
using various combinations of exclusive states
with $\hat N_B=10^9$. 
The dotted curve uses only data from two $D$ decay modes: CPES- and
$K^{*+}\pi^-$; the solid curve includes data from three modes:
CPES-, $K^{*+}\pi^-$ and $K^-\pi^+$. The dot-dashed curve uses the same
modes but correlation data from a $\psi(3770)$ factory 
with $\hat N_{D}=10^7$ 
is also included
which helps
determine the strong phase differences for each of the modes. The values
of the strong phase differences used are 
$120^\circ$ for $K^+\pi^-$ and 
$60^\circ$ for $K^{*+}\pi^-$. 
}\label{exclusives_B}
\end{figure}

In the ADS method, two modes are sufficient, in principle, to determine
$\gamma$ with some ambiguity and this is illustrated in
Fig.~(\ref{exclusives_B}). In this figure, we assume that $\hat N_B$, the
number of $B's$ times the acceptance, is $10^9$.  We assume that the
actual number of events of each type detected is equal to the theoretical
value; the horizontal axis covers various possible values of $\gamma$ and
for each hypothesized value of $\gamma$ a minimum value of $\chi^2$ is
obtained\footnote{Thus, if we assume that acceptance is 10\% and there are
$10^9$ $B$'s, a 3$\sigma$ criterion for eliminating values of $\gamma$
translates to $\chi^2>90$ on this graph.}.  The dotted curve assumes that
data is taken for just the two modes:  $B^-\to K^- [D^0\to K^{*+}\pi^-]$
and $B^-\to K^- [D^0\to {\rm CPES-}]$ (+ charge conjugates). It can be
readily seen that the ambiguities of that method tend to interfere with a
clean determination of $\gamma$.

Of course three modes will improve the situation so if we now add in the
other $K^+\pi^-$ mode, we obtain the solid curve and clearly there is some
improvement. Note that in all cases (including what we discuss below)
there is a residual 4-fold ambiguity between $\gamma$, $\pi-\gamma$,
$\pi+\gamma$ and $2\pi -\gamma$ as discussed in~\cite{ads}.

\begin{figure}
\epsfxsize 3.0 in
\mbox{\epsfbox{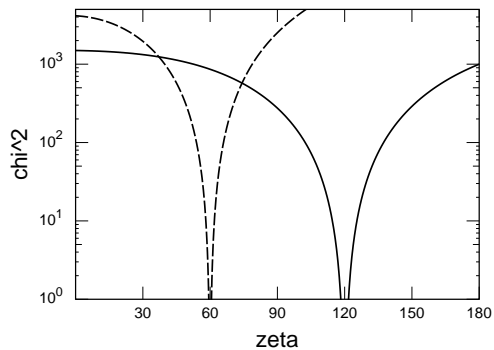}}
\caption{
$\chi^2$ is plotted as a function of $\ETA$ 
using correlations at a $\psi(3770)$ factory using 
correlations between the exclusive modes considered in
Fig.~(\ref{exclusives_B}). The solid line is for the mode $K^+\pi^-$ while
the dashed line is for the mode $K^{*+}\pi^-$.
}\label{exclusives_DD}
\end{figure}

To improve the situation still further, let us consider adding information
from a $\psi(3770)$ factory. Here we will take the number of $D^0\xba D^0$
pairs times acceptance to be $\hat N_D=10^7$. The information from the
correlations of the above two modes with each other and with the CP
eigenstates thus serve to determine $\ETA(K^+\pi^-)$ and
$\ETA(K^{*+}\pi^-)$ with a degree of over determination $\delta n=4$. in
Fig.~(\ref{exclusives_DD}) we show $\chi^2$ as a function of $\ETA$ for
these two modes. It is clear that except for the discrete ambiguities, the
values of these angles are relatively well determined.

Returning now to Fig.~(\ref{exclusives_B}) the dashed-dotted line shows
the improvement obtained by adding this information from 
the charm factory. 

Let us now consider the case where we use inclusive final states, in
particular we will consider the sets of inclusive states
$\F=\F_1\cup\F_2\cup\F_3$ discussed in our toy model together with the
CPES- states.  As discussed above, the determination of $\ETA$ and $R$
using $\psi(3770)$ factory data is shown in figures
Fig.~(\ref{inclusive_eta})  and Fig.~(\ref{inclusive_R}).

\begin{figure}
\epsfxsize 3.0 in
\mbox{\epsfbox{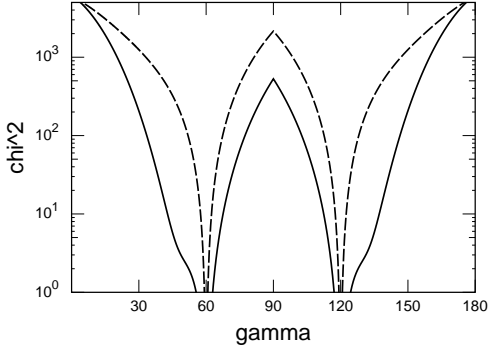}}
\caption{
$\chi^2$ is plotted as a function of $\gamma$ for input generated assuming
$\gamma=60^\circ$ and $\zeta_k=-50^\circ$ 
using various combinations of inclusive states
with $\hat N_B=10^9$ together with $\psi(3770)$ factory data.
The solid curve uses only $\F$ and CPES-. The dashed line uses information
from $\F_{1-3}$ together with CPES-.
}\label{inclusives_B}
\end{figure}

We can not use inclusive modes to find $\gamma$ via the ADS method because
of the extra degree of freedom $R$ associated with each mode. With the
data from a $\psi(3770)$ charm factory such a determination becomes
possible. This is illustrated in Fig.~(\ref{inclusives_B}). The solid
curve
shows the result obtained from $\F$ with CPES- while the dashed curve
shows the result using $\F_{1-3}$ separately with CPES-. Clearly there is
considerable gain in segregating the data into different subsets.

In Table~(\ref{tab_results}) we summarize the 3-$\sigma$ errors in the
determination of $\gamma$ which follow from the calculations in
Figs.~(\ref{exclusives_B}) and Figs.~(\ref{inclusives_B}). The
$\psi(3770)$ data clearly improves the error in the case of exclusive
states. In the case of inclusive states where $\psi(3770)$ data is
essential, we see that dividing the inclusive state into several sub modes
can lead to an improved determination of $\gamma$.

%%
%%>DA-ref[
%%
%% This improvement, in spite of the fact that each sub mode has fewer
%% events, results from two facts (1) The sub modes can have larger values
%% of
%% $R$ hence more sensitivity to $\gamma$ and (2) The greater number of
%% sub modes leads to a greater degree by which the parameters are
%% overdetermined. 
%%
%%>DA-ref]
%%

Finally, it is important to note that these methods are not limited to 
the case where the $D^0$ decay is inclusive, one can also use $R$ and 
$\zeta$ to characterize inclusive decays of the parent $B$. For instance, 
as pointed out by~\cite{bkdpi} the decay $B^-\to D^0 K^- \pi^0$ has the 
advantage that there is no color suppression in the $b\to u$ transition; 
hence larger branching ratios would be expected. Using our methods, this 
mode integrated over a portion of the $B$ Dalitz plot would be 
characterized by two parameters $R_B$ and $\zeta_B$ as opposed to just 
$\zeta_B$ for the two body $B$ decay. 

If exclusive $D$ decays are used alone then at least three modes would be 
required to determine $\gamma$. If, however exclusive and inclusive modes 
of $D^0$ decay are used where $R$ and $\zeta$ are supplied by the charm 
factory then two modes would be sufficient to determine $\gamma$. Again, 
this method makes no apriori assumptions concerning the structure of the 
$B$ decay amplitude and may be generalized to a broad class of inclusive 
decays such as $B^-\to D^0 K^-+n\pi$, $D^0 K^*\rho$ etc.

A special case of the above would be to consider the two $D^0$ decays to 
the exclusive states $K^+\pi^-$ and CPES where the $D^0$ is produced via
$B^0\to D^0K^-$, 
$B^0\to D^{*0}K^-$, 
$B^0\to D^0K^{*-}$ 
and
$B^0\to D^{*0}K^{*-}$. 
It is likely that these four modes have similar strong phases so $R_B$ due 
to the summation over these $K$ and $D$ resonances is likely to not be 
$<<1$. With a charm factory determination of $\zeta(K^+\pi^-)$ there would 
be enough information to determine $\gamma$. The advantage  is that 
all the final states here are relatively  clean experimentally and the ADS 
mode $K^+\pi^-$ 
is likely to have relatively large CP violation so this combination could 
provide an early determination of $\gamma$.

%
%----------------------------------------------------------------
%

\section{Summary}\label{summary}

In this paper we have generalized the analysis of direct CP violation from
states which are single quantum states to states which are inclusive
either because they are integrated over phase space or include states with
different particle content. In particular, we study the determination of
$\gamma$ using decays of the form $B^-\to K^- D^0$ where the $D^0$
subsequently decays to various inclusive final states. 

We have shown that 
for a given inclusive state $F$
the phase relation between $D^0\to F$ and $\xba D^0\to
F$ can be expressed in terms of a strong phase $\ETA$ and a coherence
coefficient
$R$. These quantities can be extracted at a $\psi(3770)$ charm
factory by observing correlations of the form 
$\psi(3770)\to DD\to F_iF_j$,
$\psi(3770)\to DD\to F_i\xba F_j$
and
$\psi(3770)\to DD\to F_i\ {\rm CPES}$
where $F_i$ are various inclusive states. 

This information then allows the extraction of $\gamma$ via 
CP violating effects in 
the reaction
$B^-\to K^- [D^0\to F_i]$ in a model independent way. To illustrate the
method we construct a toy model for inclusive decays of the form
$K^-+X$. We find that if this is decomposed into three subsets according
to the energy of the $K^-$, then a determination of $\gamma$ with a
3-$\sigma$ error of $2.3^\circ$ can be made if $\hat N_B=10^9$.

%
%
% \section{Extracting $\delta$ from $B^0\to D^0 K_S$}
% 
% \section{Combining Charged and Neutral Modes}
% 
% 
% \section{Conclusion}
% 
% \section*{Appendix: $D^0$ oscillation}
% 
% 
% 

\medskip

This research was supported by Contract Nos.\
DE-FG02-94ER40817 and DE-AC02-98CH10886.

%>1
%-------------------------------------->Figure 1
%

%
%-------------------------------------->Figure 1
%

%
%-------------------------------------->Figure 2
%>>2

%
%-------------------------------------->Figure 2
%

%
%-------------------------------------->final fig 3
%>>3

%
%-------------------------------------->final fig 3
%

%
%-------------------------------------->final fig 4
%>>4

%
%-------------------------------------->final fig 4
%

%
%-------------------------------------->final fig 1
%

%
%-------------------------------------->final fig 1
%

%
%-------------------------------------->final fig 2
%

%
%-------------------------------------->final fig 2
%

%
%-------------------------------------->final fig 5
%

%
%-------------------------------------->final fig 5
%

%%%%%%%%%%%%%%%%%%%%%%%%%%%%%%%%%%%%%%%%%%%%%%%%%%%%%%%%%%%%%%%%%%%
%%%%%%%%%%%%%%%%%%%%%%%%%%%%%%%%%%%%%%%%%%%%%%%%%%%%%%%%%%%%%%%%%%%
%%%%%%%%%%%%%%%%%%%%%%%%%%%%%%%%%%%%%%%%%%%%%%%%%%%%%%%%%%%%%%%%%%%
%%%%%%%%%%%%%%%%%%%%%%%%%%%%%%%%%%%%%%%%%%%%%%%%%%%%%%%%%%%%%%%%%%%
%%%%%%%%%%%%%%%%%%%%%%%%%%%%%%%%%%%%%%%%%%%%%%%%%%%%%%%%%%%%%%%%%%%
%%%%%%%%%%%%%%%%%%%%%%%%%%%%%%%%%%%%%%%%%%%%%%%%%%%%%%%%%%%%%%%%%%%
%%%%%%%%%%%%%%%%%%%%%%%%%%%%%%%%%%%%%%%%%%%%%%%%%%%%%%%%%%%%%%%%%%%
%%%%%%%%%%%%%%%%%%%%%%%%%%%%%%%%%%%%%%%%%%%%%%%%%%%%%%%%%%%%%%%%%%%
%%%%%%%%%%%%%%%%%%%%%%%%%%%%%%%%%%%%%%%%%%%%%%%%%%%%%%%%%%%%%%%%%%%
%%%%%%%%%%%%%%%%%%%%%%%%%%%%%%%%%%%%%%%%%%%%%%%%%%%%%%%%%%%%%%%%%%%
%%%%%%%%%%%%%%%%%%%%%%%%%%%%%%%%%%%%%%%%%%%%%%%%%%%%%%%%%%%%%%%%%%%
%%%%%%%%%%%%%%%%%%%%%%%%%%%%%%%%%%%%%%%%%%%%%%%%%%%%%%%%%%%%%%%%%%%
%%%%%%%%%%%%%%%%%%%%%%%%%%%%%%%%%%%%%%%%%%%%%%%%%%%%%%%%%%%%%%%%%%%
%%%%%%%%%%%%%%%%%%%%%%%%%%%%%%%%%%%%%%%%%%%%%%%%%%%%%%%%%%%%%%%%%%%

%
%------------------------------------------>
%

\begin{table}
\begin{tabular}{|c|c|c|}

\hline
Mode   &    Sub-mode   &   Br  \\
\hline
\hline
$K^-\pi^+$ &  &  3.8\% \\

\hline
$K^-\pi^+\pi^0$ &       &  13.1\% \\
& $K^- [\rho^+\to \pi^+\pi^0]$ & 8.64\% \\
& $\pi^+[K^{*-}\to K^-\pi^0]$ & 5.02\% \\
& $\pi^0 [\bar K^{*0}\to K^-\pi^0 ]$ & 1.46\%\\

\hline
$K^-\pi^-\pi^+\pi^+$&  &                         7.46\%  \\
&$K-\pi^+[\rho^0\to\pi^+\pi^-]$&                 4.7\%   \\
&$[K^{*0}\to K^-\pi^+][\rho^0\to\pi^+\pi^-]$&    0.97\%   \\
&$K^- [a_1^+\to \pi^+\pi^+\pi^-] $ &             3.6\%    \\
&$[K_1^-(1270)\to K^-\pi^+\pi^-]\pi^+ $&         0.37\%  \\
&4-body continuum &                              1.74\%   \\
\hline
$K^-\pi^-\pi^+\pi^+\pi^0$ &  & 4.0\% \\

\hline
\end{tabular}
\caption{\emph 
The modes used in the toy model for $D^0\to K^- X$ where $X$ contains
at most one $\pi^0$.
}
\label{tab1}
\end{table}

%
%------------------------------------------>
%

\begin{table}
\begin{tabular}{|c|c|c|c|}
\hline
Mode & Br  &  $\ETA$ & R \\
\hline
\hline
$\F_1 $  & 11.0\%   & $-34^\circ$    & 0.74    \\
\hline
$\F_2 $  & 24.9\%   &  $-86^\circ$   & 0.29    \\ 
\hline
$\F_3 $  & 19.3\%   &  $30^\circ$   &  0.91   \\ 
\hline
CP eigenstates  & 5\%   &     &  1   \\ 
\hline
\end{tabular}
\caption{\emph 
The parameters for the three different 
inclusive modes resulting in our toy model.
}
\label{mod_parm}
\end{table}

%
%-------------------------------------->
%
\begin{table}
\begin{tabular}{|l|c|}
\hline
Input                                           &
$\gamma=60^\circ$; $\zeta_k=-50^\circ$   
\\
\hline
$K^{*+}\pi^-$ with CPES- &
$10.0^\circ$    
\\
\hline
$K^{*+}\pi^-$ and $K^{+}\pi^-$ with CPES- &
$9.1^\circ$    
\\
\hline
$K^{*+}\pi^-$ and $K^{+}\pi^-$ with 
&
$3.4^\circ$    
\\
CPES- using $\psi(3770)$ 
&

\\
\hline
$\F$ with CPES- &
$12.0^\circ$    
\\
\hline
$\F_1$, $\F_2$ and $\F_3$ with CPES- &
$2.3^\circ$    
\\
\hline
\end{tabular}
\caption{
\emph{
The 3-$\sigma$ error in degrees in the determination of 
$\gamma$ 
with $\hat N_B=10^9$
for the various toy models considered in the text for
$\gamma=60^\circ$ and $\zeta_k=-50^\circ$. 
The first three rows refer to exclusive states where we take
$\eta(K^+\pi^-)=120^\circ$
and 
$\zeta(K^+\pi^-)=60^\circ$ corresponding to the curves in
Fig.~(\ref{exclusives_B}) while the last two rows refer to the inclusive
states $\F_i$ corresponding to the curves in Fig.~(\ref{inclusives_B})
}
}\label{tab_results}
\end{table}

%
%-------------------------------------->
%%%%%%%%%%%%%%%%%%%%%%%%%%%%%%%%%%%%%%%%%%%%%
% Additional table????
%
%
%
% \begin{table}
% \begin{tabular}{|l|c|c|}
% \hline
% Mode & Avarage Br: ${\frac12(d+\bar d)}$ 
% (10$^{-6}$)
% & CP Violation: $\left | { (d-\xba d)/
% (d +\xba d)}\right |$ (\%)
% \\
% \hline
% $K^{+}\pi^-$ & 0.55 & 57 
% \\
% \hline
% $K^{*+}\pi^-$ & 3.07 & 9 
% \\
% \hline
% CPES-    & 30.8 & 16 
% \\
% \hline
% $\F_1$   & 4.5 & 52 
% \\
% \hline
% $\F_2$  & 8.4  & 16
% \\
% \hline
% $\F_3$  & 10.6 & 16
% \\
% \hline
% \end{tabular}
% \caption{
% \emph{
% The CP violation and average brancing ratio of the modes considered in
% Table~(\ref{tab_results}) for the specific values of the parameters
% considered there 
% }
% }\label{tab_BR}
% \end{table}
%
%
%
%%%%%%%%%%%%%%%%%%%%%%%%%%%%%%%%%%%%%%%%%%%%%%


\begin{thebibliography}{99}
\bibitem{nocp} Throughout this paper Standard Model is assumed
and CP violation in $D^0- \xba D^0$ is taken to be negligibly small.
\end{thebibliography}
\end{document}